\documentstyle[epsf,psfig]{article}
\begin{document}
\centerline{\Large\bf Polarized Electromagnetic Radiation
 from }
\centerline{\Large\bf 
Spatially Correlated Sources}

\bigskip

\centerline{\bf Abhishek Agarwal\footnote{current address: Physics
Department, University of Rochester, Rochester, NY}, 
Pankaj Jain\footnote{e-mail: pkjain@iitk.ac.in} and Jagdish Rai}

\begin{center}
Physics Department\\
Indian Institute of Technology\\
Kanpur, India 208016\\

\end{center}

\begin{abstract}

We consider the effect of spatial correlations on sources of 
polarized electromagnetic radiation.  
The sources, assumed to be monochromatic, are constructed out of
dipoles aligned along a line such that their orientation
is correlated with their position. 
In one representative example, the dipole orientations 
are prescribed by a generalized form of the standard von Mises 
distribution for angular variables such that the azimuthal angle
of dipoles is correlated with their position. In another example
the tip of the dipole vector
traces a helix around the symmetry axis of the source, thereby
modelling the DNA molecule.  
We study the polarization properties
of the radiation emitted from such sources in the radiation zone.
For certain
ranges of the parameters we find a rather striking angular dependence
of polarization. This may find useful applications in certain biological
systems as well as in astrophysical sources. 

\end{abstract}

\section{Motivation}

In a series of interesting papers Wolf  
\cite{Wolf86,Wolf87,Wolf87A}
studied the spectrum of light
from spatially correlated sources and found, remarkably, that in general
the spectrum
does not remain invariant under propagation even through vacuum.
The phenomenon was later confirmed experimentally 
\cite{Faklis88,Gori88,Inde89} and has been a subject of considerable 
interest \cite{review}. Further investigations of the source correlation
effects have been done in the time domain theoretically \cite{Rai}
and experimentally \cite{Chopra}.
Several applications of the effect have also
been proposed
\cite{James95,Kandpal95,Vicalvi96,Wolf97,Shirai98}. In a related
development it has been pointed out that spectral changes also 
arise due to static scattering \cite{Foley89,Shirai95,Shirai96,Leskova97,Dogriu} and 
dynamic scattering \cite{Wolf89,Foley89A,James90,JW90,JW94}.
In the present paper we investigate 
polarization properties of spatially correlated sources.
Just as we expect spectral shifts for such 
sources, we expect nontrivial polarization effects if the correlated 
source emits polarized light.   

The basic idea can be illustrated by considering 
two polarized point sources $P_1$ and 
$P_2$ which are located along the z-axis at a distance $2z$ 
apart, in analogy to a similar situation considered by Wolf \cite{Wolf87}
for the unpolarized case. The electric field at the point $Q$ located
at large distances $R_1$ and $R_2$ from these point sources can be written as,
\begin{equation}
\vec E = \vec e(\vec r_1,\omega) {e^{ikR_1}\over R_1} + \vec e(\vec r_2,\omega) 
{e^{ikR_2}\over R_2}
\end{equation}
Here $\vec r_1$ and $\vec r_2$ are the 
position vectors of the points $P_1$ and $P_2$.
We calculate the coherency matrix for this electric field at the point Q,
\begin{eqnarray}
<E^*_iE_j> &=& {<e^*_i(\vec r_1,\omega)e_j(\vec r_1,\omega)>\over R_1^2} + 
{<e^*_i(\vec r_2,\omega)e_j(\vec r_2,\omega)>\over R_2^2}\nonumber\\ 
&+& 
\left({<e^*_i(\vec r_1,\omega)e_j(\vec r_2,\omega)>e^{ik(R_2-R_1)}\over 
R_1R_2} + h.c.\right)
\end{eqnarray}
We are interested in investigating the effect of the third term which arises due
to cross correlation between the sources $P_1$ and $P_2$. It is clear that 
this term will give nontrivial contribution to the polarization in the far zone.
In the present paper we investigate the contribution of this 
term by considering only 
monochromatic waves. It turns out to be useful to understand this simple
idealization before treating the realistic case of 
partially polarized radiation. 

In order to illustrate the contribution of the cross correlation term
we consider the situation where the two sources are
simple dipoles, $\vec p_1$ and $\vec p_2$ located
at $z$ and $-z$ respectively and are oriented
such that their polar angles $\theta_1=\theta_2=\theta_p$ and the
azimuthal angles $\phi_1=-\phi_2=\pi/2$. We assume that $\theta_p$
lies between 0 and $\pi/2$. The strength of the dipoles
is $p_0$ and they radiate at frequency $\omega$.
We compute the electric field at point $Q$ located 
at coordinates $(R,\theta,\phi)$, 
as shown in Fig. 1, 
such that $R>>z$. The electric field in the far zone
is obtained by the addition of two vectors $\hat
p_1\cdot \hat R \hat R - \hat p_1$
and $\hat p_2\cdot \hat R \hat R - \hat p_2$ with phase difference of
$2\omega z\cos\theta/c$. 
The
vector $\hat p\cdot \hat R \hat R - \hat p$ at any point $(R,\theta,\phi)$
is ofcourse simply the 
projection of the polarization vector
$\hat p$ on the plane perpendicular to $\hat R$ at that point.  
We keep only the leading order terms in $z/R$ in computing the total electric
field.

The observed polarization is obtained by calculating the coherency
matrix, given by
\begin{equation}
J=\left( \begin{array}{cc}\langle E_{\theta }E_{\theta }^* \rangle &
\langle E_{\theta }E_{\phi }^* \rangle \\ \langle E_{\phi }^* E_{\theta
}\rangle & \langle E_{\phi }E_{\phi }^* \rangle \end{array}\right)
\end{equation}
The state of polarization can be uniquely specified by the Stokes's parameters
or equivalently the Poincare sphere variables \cite{Born}. 
The Stoke's parameters are obtained in
terms of the elements of the coherency matrix as:
$S_0 = J_{11}+J_{22},\ \ 
S_1 = J_{11}-J_{22},\ \ 
S_2 = J_{12} + J_{21},\ \ 
S_3 = i(J_{21} - J_{12})$.
The parameter $S_0$ is proportional to the intensity of the beam.
The Poincare sphere is charted by the angular variables $2\chi $, and $2\psi
$, which can be expressed as:
\begin{equation}
S_1 = S_0\cos2\chi \cos 2\psi,\ \ 
S_2 = S_0\cos2\chi \sin 2\psi,\ \ 
S_3 = S_0\sin2\chi 
\end{equation}

The angle $\chi$ ($-\pi/4\le \chi\le \pi/4$) 
 measures of the ellipticity of the state of polarization
and $\psi$ ($0\le \psi< \pi$) 
measures alignment of the linear polarization. For example,
$\chi=0$ represents pure linear polarization and $\chi=\pi/4$ pure right 
circular 
polarization.   

The Stokes parameters at the observation point $Q$ for the two dipole source
are given by
$$ S_0 = \left({\omega^2p_0\over c^2 R}\right)^2\left[4\cos^2(\omega z\cos
\theta/c)\cos^2\theta_p\sin^2\theta+4\sin^2(\omega z\cos\theta/c)\sin^2\theta_p
(\cos^2\theta\sin^2\phi+\cos^2\phi)\right] $$ 
$$ S_1 = \left({\omega^2p_0\over c^2 R}\right)^2\left[4\cos^2(\omega z\cos
\theta/c)\cos^2\theta_p\sin^2\theta+4\sin^2(\omega z\cos\theta/c)\sin^2\theta_p
(\cos^2\theta\sin^2\phi-\cos^2\phi)\right] $$ 
$$ S_2 = \left({\omega^2p_0\over c^2 R}\right)^28\sin^2(\omega z\cos
\theta/c)\sin^2\theta_p\cos\theta\sin\phi\cos\phi
$$
$$ S_3 = \left({\omega^2p_0\over c^2 R}\right)^28
\cos(\omega z\cos\theta/c)\sin(\omega z\cos
\theta/c)\cos\theta_p\sin\theta_p\sin\theta\cos\phi$$

The resulting polarization in the far zone is quite interesting. At 
$\sin\phi=0$, $\theta=\pi/2$ the wave is linearly polarized ($\chi=0$) 
with $\psi=0$. As $\theta$ decreases from $\pi/2$ to $0$, $\chi>0$ and 
the wave has general elliptical polarization with $\psi=0$. At 
a certain value of the polar angle $\theta=\theta_t$
the wave is purely right circularly
polarized. As  $\theta$ crosses $\theta_t$, the linearly polarized component
jumps from $0$ to $\pi/2$, i.e. $2\psi$ changes from 0 to $\pi$. 
The value of the polar angle $\theta_t$
at which the transition occurs is determined
by
\begin{equation}
\tan(\omega z\cos\theta_t/c) = \pm \sin\theta_t/\tan\theta_p
\end{equation}
From this equation we see that as $z\rightarrow 0$, the transition angle 
$\theta_t$ is close to
zero for a wide range of values of
$\theta_p$. Only when $\theta_p\rightarrow \pi/2$, a solution with
$\theta_t$ significantly different from
0 can be found. In general, however, we can find a solution with any value
of $\theta_t$   by appropriately 
adjusting $z$  and $\theta_p$. 
For $\sin\phi>0$, we find qualitatively similar results, except that 
now the linear polarization angle $2\psi$ increases smoothly from $0$
as $\theta$ goes from $\pi/2$ to $0$. Furthermore the state of
polarization never becomes purely circular, i.e. the angle $2\chi$ 
is never equal to $\pi/2$ although it does achieve a maxima at
$\theta$ close to the transition angle $\theta_t$ given by Eq. 5.


We therefore find that for wide range of parameters the polarization
in the far field region shows a nontrivial dependence on angular position
which is governed by the cross correlation term. 
In general this term
will give nontrivial contribution unless the two sources have identical
polarizations. 

\section{A One Dimensional Spatially Correlated Source}
A simple continuous model of a monochromatic spatially correlated 
source can be constructed by arranging a series of dipoles along a line 
with their 
orientations correlated with the position of the source. The dipoles 
will be taken to be
aligned along the $z$ axis and distributed as a gaussian
$\exp[-z^2/2\sigma^2]$. The orientation of the dipole is characterized
by the polar coordinates $\theta_p,\phi_p$, which are also assumed
to be correlated with the position $z$. A simple correlated ansatz is
given by
\begin{equation} 
{\exp\left[\alpha\cos(\theta_p) + \beta z\sin(\phi_p)\right]\over
N_1(\alpha)N_2(\beta z)} 
\end{equation} 
 where $\alpha$ and $\beta$ are parameters, 
$N_1(\alpha) = \pi I_0(\alpha)$ and $N_2(\beta z) = 
 2\pi I_0(\beta z)$ are normalization factors and $I_0$ is the Bessel
function. The basic distribution function $\exp(\alpha\cos(\theta-\theta_0))$
used
in the above ansatz is the well known von Mises distribution which
for circular data is in many ways the analoque of Gaussian distribution for
linear data \cite{Mardia,Batschelet,Fisher}. 
For $\alpha>0$ this function peaks at
$\theta=\theta_0$. Making a Taylor expansion close to its peak we find a 
gaussian distribution to leading power in $\theta-\theta_0$. The maximum
likelihood estimators for the mean angle $\theta_0$ and the width
parameter $\alpha$ are given by, 
$<\sin(\theta-\theta_0)>=0$ and $<\cos(\theta-\theta_0)> = d \log (I_0(\alpha))
/ d\alpha$ 
respectively. In prescribing the ansatz given in Eq. 6 we have assumed that
the polar angle $\theta_p$
of the dipole orientation is uncorrelated with $z$ and the distribution
is peaked either at $\theta_p=0$  ($\pi$) for $\alpha>0$ ($<0$). 
The azimuthal angle $\phi_p$ is correlated with $z$ such
that for $\beta>0$ 
and $z>0(<0)$ the distribution peaks
at $\phi_p=\pi/2(3\pi/2)$.     

\begin{figure}
\raisebox{-1.1in}{\psfig{file=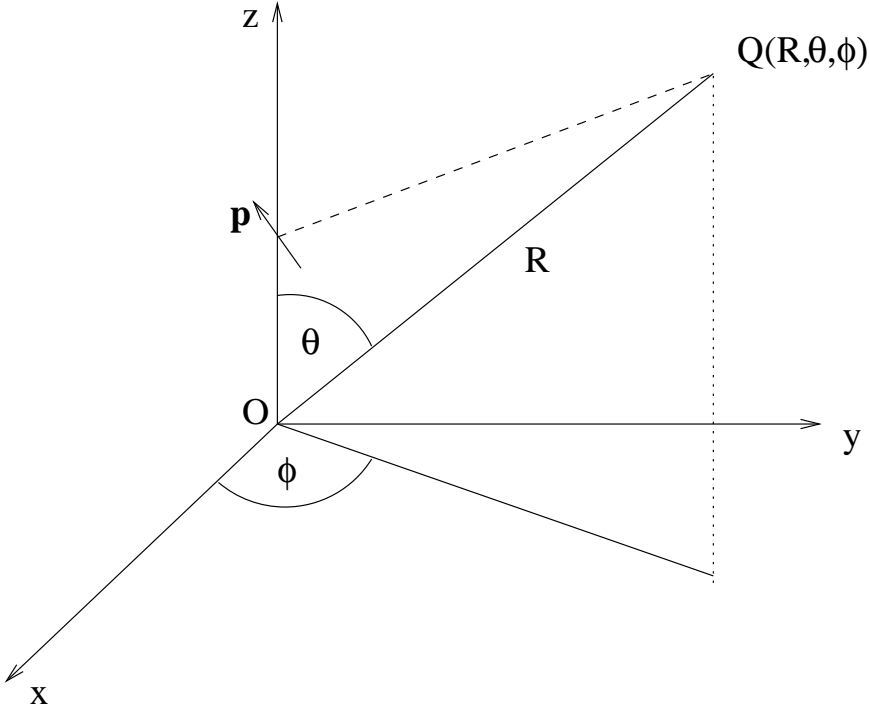,width=0.7\textwidth}}
\caption{The correlated source consisting of an array of dipoles
aligned along the $z$ axis. The observation point $Q$ is at a 
distance $R$ which is much larger than the spatial extent of the
source.}
\end{figure}

We next calculate the electric field at very large distance from 
such a correlated source. The observation point $Q$ is located at
the position $(R,\theta,\phi)$ (Fig 1), measured in terms of the spherical 
polar coordinates, and we assume that the spatial
extent of the source $\sigma<<R$.  
The electric field  
from such a correlated source at large distances
  is given by,
\begin{eqnarray}
 E & =&  -{\omega^2\over c^2 R}p_0e^{i\left(-\omega t + R\omega/c\right)}
\int_{-\infty}^{\infty} dz \exp\left(-z^2/2\sigma^2\right)
\int_0^\pi d\theta_p\int_0^{2\pi} d\phi_p \cr
& \times& 
{\exp\left(\alpha\cos\theta_p+\beta z\sin\phi_p\right)\over 2\pi^2 I_0(\alpha)
I_0(\beta z)}
 \exp{\left(i\omega z \hat R\cdot\hat z/c\right)}
\times  \large(\hat p\cdot \hat R \hat R - \hat p\large) 
\end{eqnarray} 
where $\hat p$ is a unit vector parallel to the dipole axis, $p_0$ is
the strength of the dipole,  $\omega$ is the frequency of light and $I_0$
denotes the Bessel function. Since we are interested in the radiation zone 
we have dropped all terms higher order in $z/R$. The resulting field is
ofcourse transverse i.e. $\vec E\cdot \hat R=0$. 
We have also assumed that all the dipoles radiate at same frequency 
and are in phase. The correlation of the source with position is measured
by the parameter $\beta$. 

It is convenient to define scaled variable $\overline z = z/\sigma$,
$\overline \lambda=\lambda/\sigma$ where $\lambda=2\pi\omega/c$ 
is the wavelength, and $\overline \beta=\beta\sigma$.
The integrations over
$\theta_p$ and $\phi_p$ can be performed analytically. We numerically
integrate over $z$ for various values of position of the observation
point, the parameter $\alpha$ which determines the width of the distribution
of $\theta_p$ and for different value of the correlation parameter $\beta$.

We first study the situation where
$\overline\beta> 0$ and $\alpha>0$. 
The result for several values of $(\theta,\phi)$ are given in
figures 2,3  which show plots of the Poincare sphere variables
$2\chi$ and $2\psi$.   
The scaled wavelength $\overline\lambda=\lambda/\sigma$ 
of the emitted radiation is taken to be 
equal to $\pi$, i.e. the effective size of the source $\sigma$
is of the order of the wavelength $\lambda$. 
The results show several interesting aspects. The ellipticity 
of the state of polarization shows significant dependence on the
position of the observer. The angle $\chi=0$, i.e. the beam has pure
linear polarization, for the polar angle 
$\cos(\theta)=0,1$ for
all values of azimuthal angle $\phi$. 
It deviates significantly from 0 as $\cos(\theta)$ varies from 0 
to 1. For $\sin(\phi)=0$, $2\chi=\pi/2$ at some critical value
$\theta_t$ as  $\cos(\theta)$ varies between $0$
and $\pi/2$, i.e. the state of polarization is purely right 
circular at $\theta=\theta_t$. 
For $\sin(\phi)>0$, $2 \chi$ also deviates significantly from
0 and displays a peak at some  value of $\theta$. The precise position
of the peak is determined by the values of the correlation parameters
 $\alpha$ and $\overline\beta$. 

The alignment of linear polarization also shows some very interesting
aspects. For $\sin(\phi)=0$, we find that $\psi$ is either $0$ or 
$\pi$ depending on the value of $\theta$. The transition occurs at
the same critical value of $\theta$ where the angle $\chi$ shows a 
peak. The state of polarization is purely linear with the electric field
along the $\hat\theta$ for $\cos(\theta)=0$ and then
acquires a circular component for increasing values of $\cos(\theta)$.
At the transition point $\theta=\theta_t$, the polarization is purely
circular.
With further increase in value of $\theta$ the state of 
polarization is elliptical with the linearly polarized component 
aligned along $\hat\phi$. 
The transition point is clearly determined by the condition
$S_1 = J_{11}-J_{22}=0$. 

For other values of $\sin(\phi)$ we find $\psi=0$ for 
$\cos\theta=0$ and then deviates significantly from 0 as $\theta$\
approached $\theta_t$, finally levelling off as $\cos\theta$ 
approches 1. The final value of $\psi$ at $\cos\theta=1$ depends on the
correlation parameters and $\sin\phi$ but for a wide range of 
parameters $2\psi>\pi/2$. 
We see from Fig. 3 that the linear polarizations from
sources of this type shows striking characteristic, i.e. that the 
polarization angle $\psi$ is either close to 0 or $\pi/2$ depending on
the angle at which it is viewed.    

For $\sin\phi<0$ the Poincare sphere polar angle $2\chi$ is same as 
for $\sin\phi>0$, however the orientation of the linear polarization
$2\psi$ lies between $\pi$ and $2\pi$, i.e. in the third and fourth
quadrants of the equatorial plane on the Poincare sphere. 
For a particular value of $\phi$ the azimuthal  
angle $\psi(\phi) = - \psi(-\phi)$.

If we change the sign of $\alpha$ we do not find any change in linear
polarization angle $\psi$ however the value of $\chi$ changes sign, i.e.
the state of polarization changes from right elliptical to left elliptical.
Change in sign of $\overline\beta$ also 
leaves $\psi$ unchanged while changing the
sign of $\chi$. 
Changing the signs of both $\alpha$ and $\overline\beta$ produces
no change at all. 

In the case of the limiting situation where 
$\overline\beta=0$
we find, as expected, 
linear polarization
is independent of the angular position, i.e. $\chi=0$
and $\psi=0$. This is true for any value of 
the parameter $\alpha$, which determines the
polar distribution of the dipole orientations. 
Hence we see that the effect disappears if either the effective size
of the source $\sigma=0$ or the correlation parameter $\beta=0$.
The effect also dissappears in the limit $\alpha \rightarrow \infty$.
In this limit the distribution of $\theta_p$ is simply a delta function
peaked at $0$ and hence our model reduces to a series of dipoles
aligned along the z-axis, which cannot give rise to any nontrivial 
structure. 
In the numerical calculations above we have taken the effective size
of the source $\sigma$ of the order of the wavelength $\lambda$.
If the size $\sigma<<\lambda$,
the effect is again 
negligible since the phase factor $\omega z\hat R\cdot\hat z/c
$ in Eq. 7 is much smaller than one in this case.

Hence we find that in order to obtain a nontrivial angular dependence
of the state of polarization the size of the source, assumed to be 
coherent, has to be of the order of or larger than the wavelength as well 
as the correlation length $1/\beta$. 

\subsection{Transition angle}
From our results we see that there exists a critical value of the
polar angle $\theta$ at
which the state of linear polarization changes very rapidly. This is
particularly true if we set $\sin\phi=0$ where we find that  
that the orientation
of linear polarization $\psi$ suddenly jumps from $0$ (or $\pi$) to
$\pi/2$ at some critical value of the polar angle $\theta=\theta_t$.
We study this case in a little more detail. 
The $\theta$ and $\phi$ components of the total electric field is given by,
$$E_{\theta} = -{\omega^2 p_0\over c^2 R} e^{-i\omega(t-R/c)}\sqrt{2\sigma^2\pi}
e^{-\sigma^2\omega^2\cos^2\theta/2c^2}{I_1(\alpha)\over I_0(\alpha)}\sin\theta$$

$$E_{\phi} = i{\omega^2 p_0\over c^2 R} e^{-i\omega(t-R/c)}
{2\sinh\alpha\over \alpha\pi I_0(\alpha)}A$$
$$A = \int_{-\infty}^\infty dz e^{-z^2/2\sigma^2} \sin(\omega z\cos\theta/c)
{I_1(\beta z) \over I_0(\beta z)} $$
In this case the Stokes parameter $S_2=0$. For $\beta\ge(<) 0$, 
$S_3 \ge (<) 0$ and hence $\chi \ge (<)0$.
The point
where the polarization angle $2\psi$ jumps from $0$ to $\pi$ is
determined by the condition $S_1=0$. This is clearly also the point
where $2\chi=\pm \pi/2$. Explicitly
the condition to determine the critical value $\theta_t$ is,
$$A^2 = \sigma^2\pi^3\alpha^2 e^{-\sigma^2\omega^2\cos^2
\theta_t/c^2}\sin^2\theta_t 
{I_1(\alpha)^2\over 2\sinh^2\alpha}.
$$
This can be used to determine $\theta_t$ as a function of $\alpha,\beta$.
The result for $\cos\theta_t$ as a function of $\alpha$
is plotted in figure 4 for several different values of $\beta$.
For any fixed value of the parameter $\beta$, the transition angle 
$\theta_t$ decreases from $\pi/2$ to 0 as $\alpha$ goes from zero to
infinity. This is expected since as $\alpha$ becomes large the polar
angle distribution of the dipole orientations, peaked along the 
$z$ axis, becomes very narrow
and hence the resultant electric field is aligned along the $z$ axis for 
a large range of polar angle $\theta$.    
Furthermore we find, as expected, that as $\beta$ goes to zero the
transition angle also tends towards 0.

\section{Helical Model}
We next study an interesting generalization of the model discussed above.
Instead of the having
the peak of the 
$\phi_p$ distribution fixed to $-\pi/2$ for $z<0$ and $\pi/2$ for $
z>0$ we allow it to rotate in 
a helix circling around the z-axis. In this case we replace the
$\phi_p$ dependence by $\exp[\beta(\phi_p-\xi z)]$. As z goes from
negative to positive values, the peak of the distribution rotates 
clockwise around the z-axis forming an helix. This is a reasonable model 
of the structure
of DNA molecule and hence has direct physical applications. 
We study this in detail
by fixing the azimuthal angle of the dipole orientation
$\phi_p=\xi z$ and the polar angle $\theta_p$ to some constant value,
i.e. the $\phi_p$ and $\theta_p$ distributions are both assumed
to be delta functions.
This allows us to perform the  $z$ integration in Eq. 7 analytically.
The resulting state of polarization, described by Poincare
sphere angles $2\chi$ and $2\psi$ are shown in Figs. 5-8 . 
In this model we can extract a simple rule to determine the transition
angle for the special case $\theta_p=\pi/2$ and $\xi=n\pi$ where
$n$ is an integer. We set $\sin\phi=0$ for this calculation 
since it is only for this value that
the polarization becomes purely circular for some value of $\theta=\theta_t$
and the linearly polarized component flips by $\pi/2$ 
at this point. A straightforward calculation shows that this
transition angle $\theta_t$ is given by:
$$\cos^2\theta_t = n\lambda/2$$
Here $n$ represents the number of $\pi$ radians that are traversed by the
tip of electric field vector along the helical path and $\lambda$ is the
wavelength. In order to get at least one transition $\lambda <2/n$.
In the special case under consideration there is atmost one transition.
However in general the situation is more complicated and for certain
values of $\theta_p$ and $\xi$, more than one transitions are possible.
Some representative examples are shown in Figs. 5-8.

\section{Conclusions}
In this paper we have considered spatially correlated monochromatic sources.
We find that at large distance the polarization of the wave shows 
dramatic dependence on the angular position of the observer. 
For certain set of parameters the linearly polarized component
shows a sudden jump by $\pi/2$. If the symmetry axis of the source is
taken to be the z-axis, the polarization shows a sudden transition from 
being parallel to perpendicular
to the symmetry axis of the source, as the polar angle is changed from
$\pi/2$ to 0. The sources considered in this paper are idealized since
we have assumed coherence over the entire source. 
For small enough sources,
such as the DNA molecule, this may a reasonable approximation. 
In the case of  macroscopic sources,
this assumption is in general not applicable.
However in certain situations some aspects of the behavior
described in this paper may survive even for these cases. For example,
we may consider a macroscopic source consisting of large number of structures
of the type considered in this paper.  As long as there is some correlation
between the orientation of these structures over large distances we expect 
that some aspects of the angular dependence of the polarization of the
small structures will survive, even if there does not exist any coherent phase
relationship over large distances. Hence the ideas discussed in this paper
may also find interesting applications to macroscopic and  
astrophysical
sources. 
As an interesting example we consider
astrophysical sources of radio waves. It is well known that the
polarization angle of these sources is predominantly observed to be
aligned either parallel or perpendicular to the source orientation axis
\cite{Clark}. This difference has generally been 
attributed to the existence of different
physical mechanism for the generation of radio waves in these sources. 
Our study, however, indicates that this difference in observed polarization
angle could also arise simply due to different angles of observation.
Hence orientation effects must be considered before attributing 
different physical mechanisms for
differences in observed polarizations of these sources.  

\bigskip
\noindent
{\bf Acknowledgements:} We thank John Ralston for very useful comments.
This work was supported in part by a research  
grant from the Department of Science and
Technology.

\begin{figure}
\psfig{file=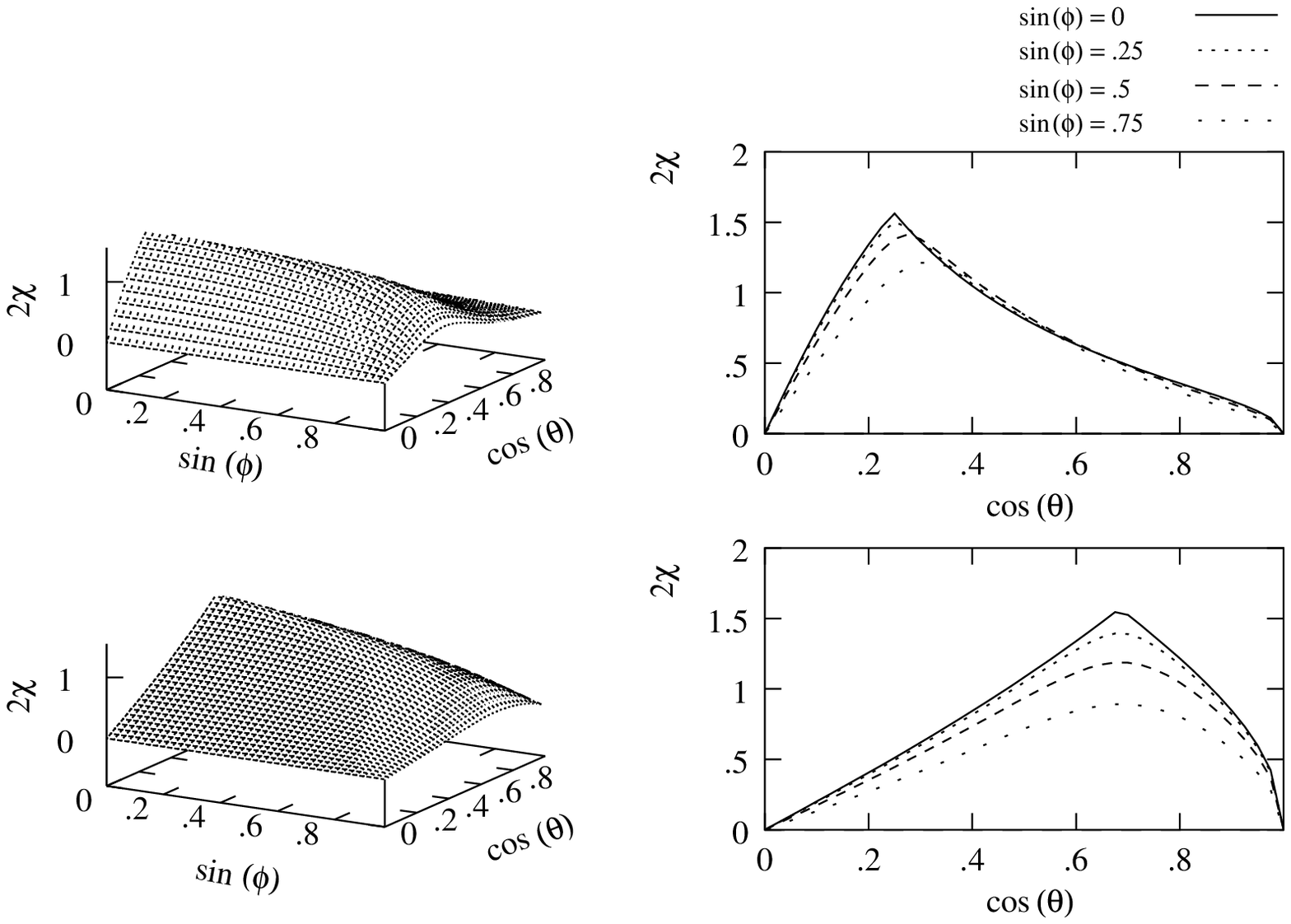}
\caption{The polar angle on the Poincare sphere $2\chi$, which is
a measure of the eccentricity of the ellipse traced by the electric
field vector. For pure linear polarization $2\chi=0$ and for pure
right circular polarization $2\chi=\pi/2$. 
The 3-D plot shows $2\chi$
as a function of $\cos\theta$ and $\sin\psi$ where $\theta$ and $\phi$
are the polar and azimuthal angles of the point of observation. The
2-D plots on the right show the corresponding slices of the 3-D plots
for different values of $\sin\phi$. 
The upper and lower plots correspond to 
$\overline\beta=1$, $\alpha=0.25$ and $\beta=\alpha=1$ respectively. 
 }
\end{figure}

\bigskip
\begin{figure}
\psfig{file=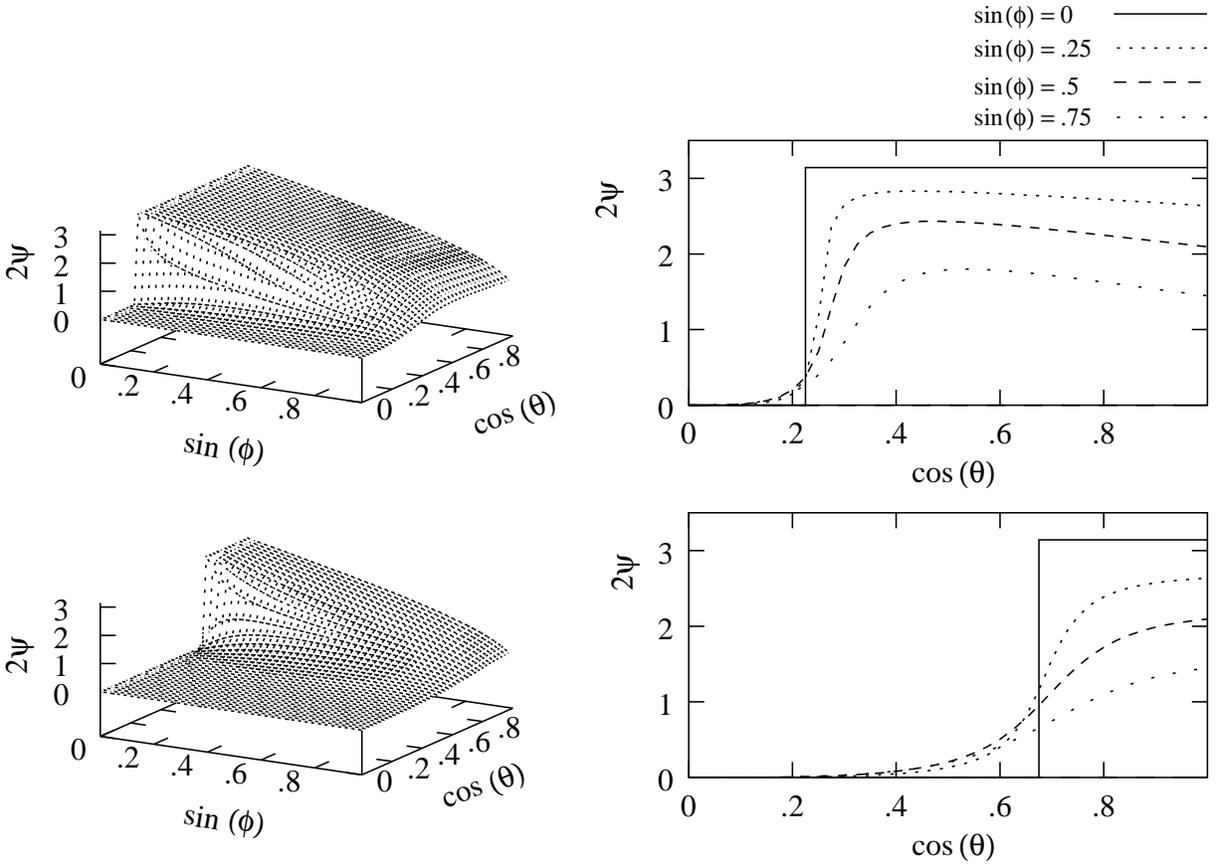}
\caption{The azimuthal angle on the Poincare sphere $2\psi$. This measures
the orientation of the linearly polarized component of the wave.
The 3-D plot shows $2\psi$
as a function of $\cos\theta$ and $\sin\psi$ where $\theta$ and $\phi$
are the polar and azimuthal angles of the point of observation. The
2-D plots on the right show the corresponding slices of the 3-D plots
for different values of $\sin\phi$. 
The upper and lower plots correspond to 
$\overline\beta=1$, $\alpha=0.25$ and $\beta=\alpha=1$ respectively. 
}
\end{figure}

\bigskip
\begin{figure}
\psfig{file=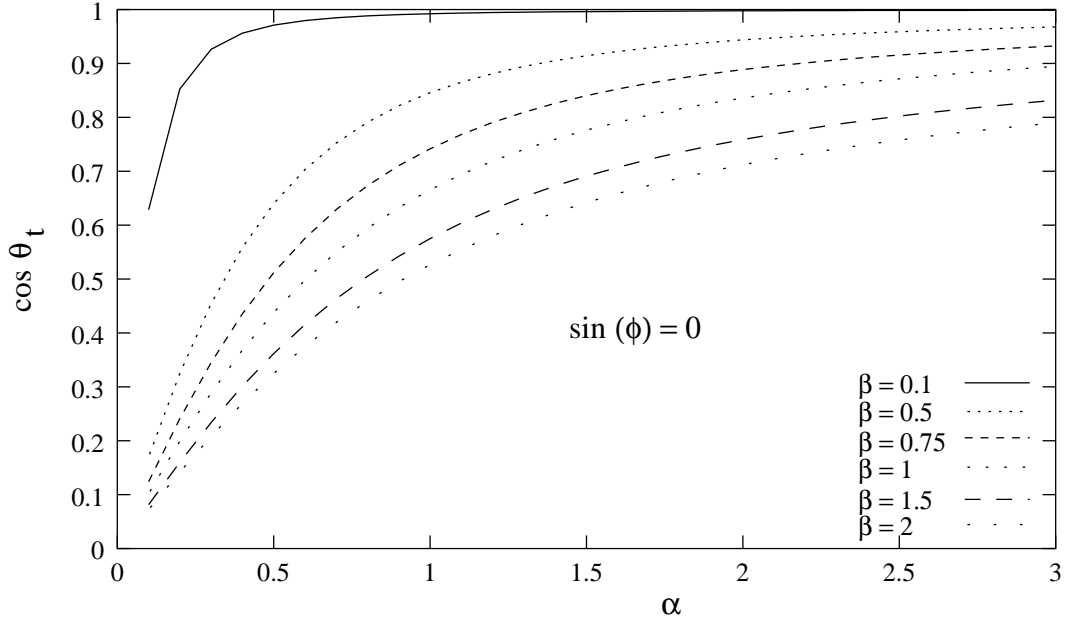}
\caption{The critical value of the polar angle $\theta$ 
 at which
the state of linear polarization shows a sudden transition for $\sin\phi=0$ 
as a function of the parameters $\alpha$ and $\beta$ which 
specify the distribution of the dipole orientations. For any
given value of the parameters $\alpha$ and $\beta$,
electric field is parallel $(\psi=0)$  
to $z$ axis if the cosine of the observation polar angle $\cos\theta$
is less than $\cos\theta_t$. On the other hand electric field is 
perpendicular to the $z$ axis if $\cos\theta$ is greater than $\cos\theta_t$. 
 }
\end{figure}

\bigskip
\begin{figure}
\psfig{file=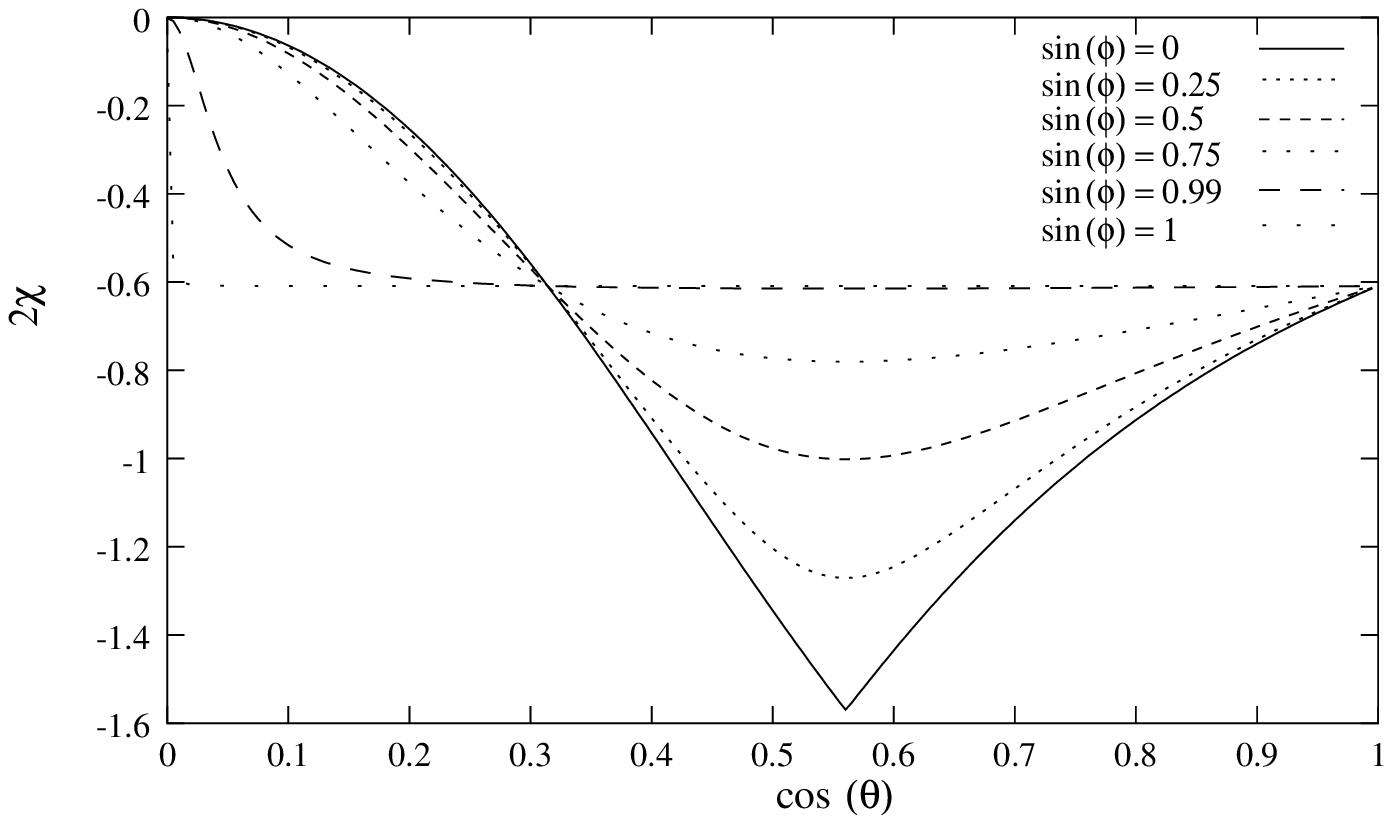}
\caption{The polar angle on the Poincare sphere $2\chi$ (radians) for the
helical model as a function of $\cos(\theta)$ ($\lambda=0.2\pi,\theta_p=\pi/2,
\xi = \pi$). 
}
\end{figure}

\bigskip
\begin{figure}
\psfig{file=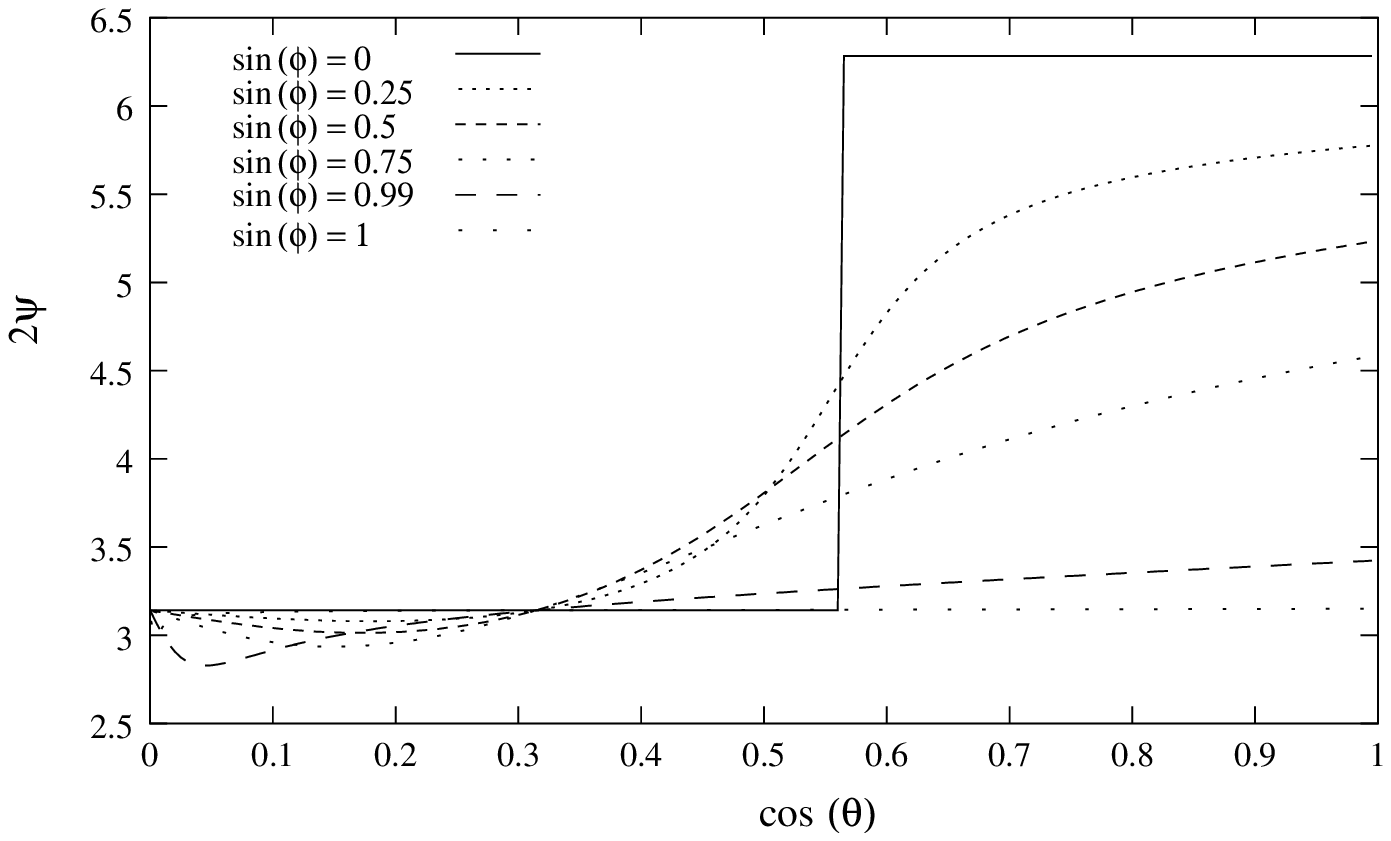}
\caption{The azimuthal angle on the Poincare sphere $2\psi$ (radians) for the
helical model as a function of $\cos(\theta)$ ($\lambda=0.2\pi,\theta_p=\pi/2,
\xi = \pi$).
}
\end{figure}

\bigskip
\begin{figure}
\psfig{file=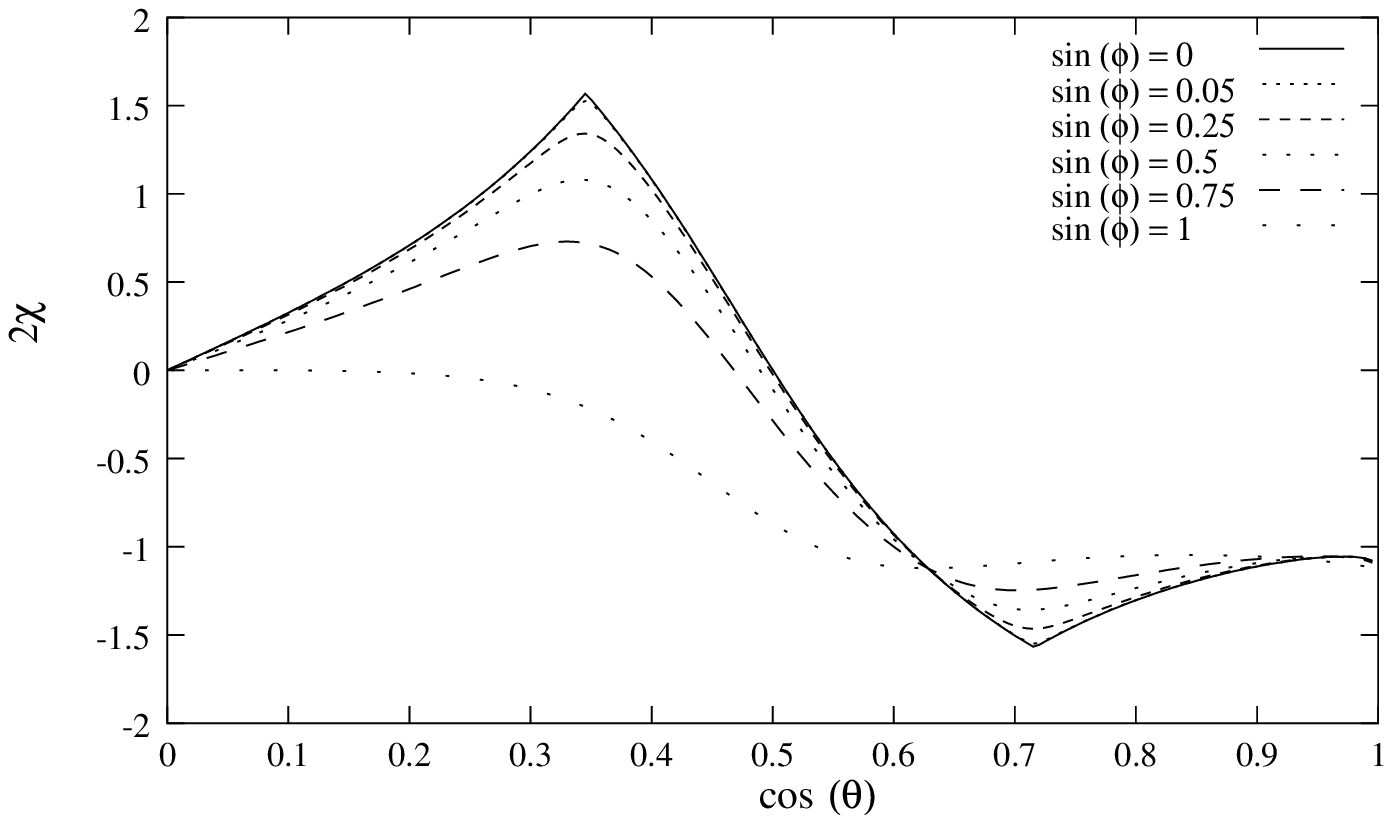}
\caption{The polar angle on the Poincare sphere $2\chi$ (radians) for the
helical model as a function of $\cos(\theta)$ ($\lambda=0.4\pi,\theta_p=\pi/4,
\xi = \pi$).
}
\end{figure}

\bigskip
\begin{figure}
\psfig{file=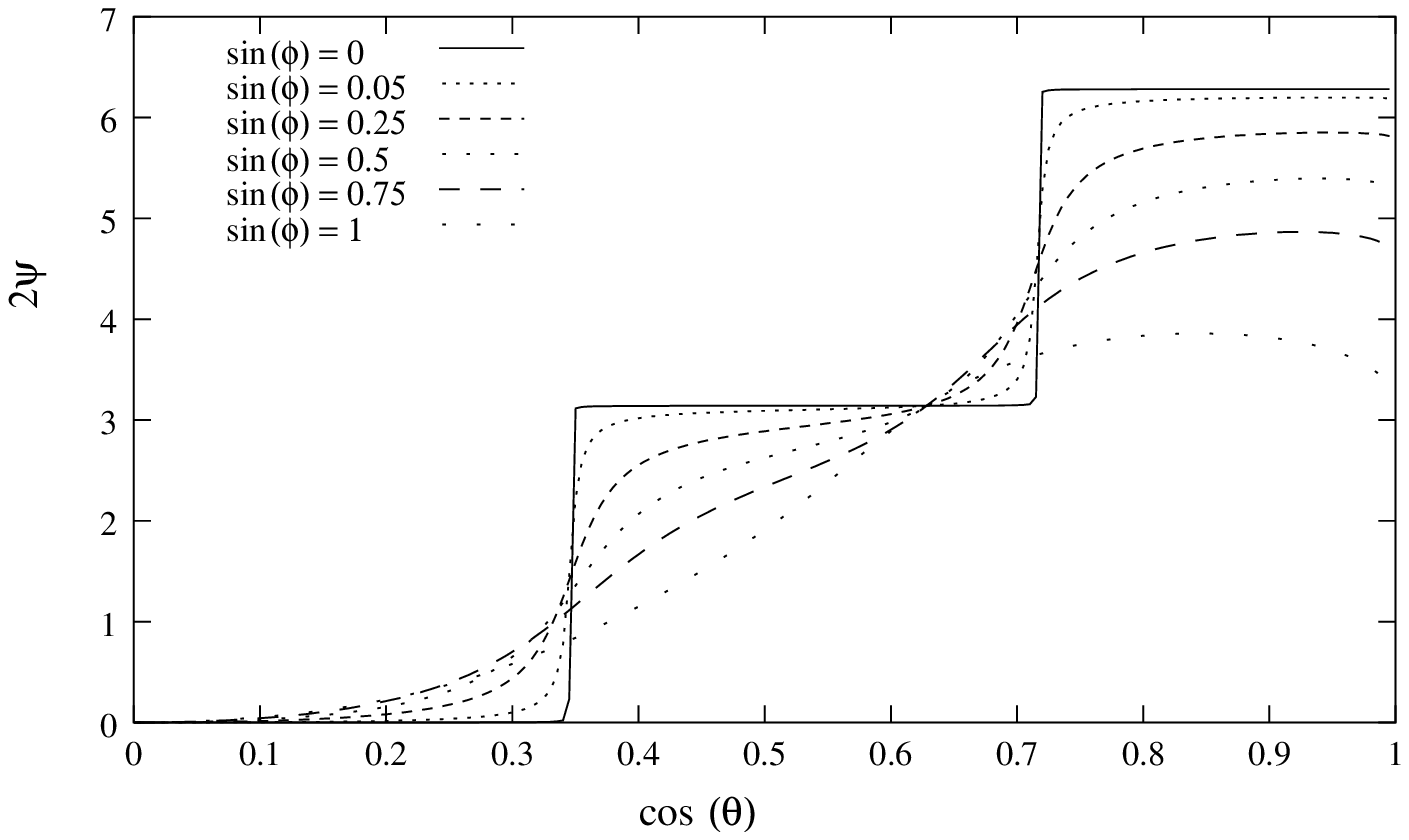}
\caption{The azimuthal angle on the Poincare sphere $2\psi$ (radians) for the
helical model as a function of $\cos(\theta)$ ($\lambda=0.4\pi,\theta_p=\pi/4,
\xi = \pi$).
}
\end{figure}

\end{document}